# Ada-MAC: An Adaptive MAC Protocol for Real-time and Reliable Health Monitoring

Feng Xia[1], Linqiang Wang[1], Daqiang Zhang[2], Xue Zhang[1], Ruixia Gao[1]

[1]School of Software, Dalian University of Technology, Dalian 116620, China

[2]School of Computer Science, Nanjing Normal University, Nanjing 210097, China

f.xia@ieee.org

*Abstract*—**IEEE 802.15.4 is regarded as one of the most suitable communication protocols for cyber-physical applications of wireless sensor and actuator networks. This is because this protocol is able to achieve low-power and low-cost transmission in wireless personal area networks. But most cyber-physical systems (CPSs) require a degree of real-time and reliability from the underlying communication protocol. Some of them are stricter than the others. However, IEEE 802.15.4 protocol cannot provide reliability and real-time transmission for time-critical and delay-sensitive data in cyber-physical applications. To solve this problem, we propose a new MAC protocol, i.e. the Ada-MAC protocol, which is based on IEEE 802.15.4 beacon-enabled mode. It can support cyber-physical applications such as health monitoring, which require stringent real- time and reliability guarantees. We implement the proposed protocol on the OMNET++ platform and conduct a performance evaluation of the proposed protocol with comparison against the traditional IEEE 802.15.4 protocol. The results are presented and analyzed.**

## I. Introduction

In recent years, wireless sensor networks (WSNs) have been applied to more and more application domains such as health monitoring, building automation, environment monitoring and industrial control. Many of these cyber-physical systems (CPSs) require a real-time and reliable communication protocol [1-5]. Health monitoring systems are a typical example. In these systems, due to some important data related to people's safety, these important data need to be delivered from the source node to the destination node in a certain time correctly in order to guarantee the validity of the data. Here, this certain time can be called the deadline, which is the maximum time which one transmission must complete its execution within. In such kind of cases, it is very necessary to employ a real-time and reliable communication protocol.

Due to its special focus on low-cost and low-power, the IEEE 802.15.4 protocol is used in WSNs widely. Meanwhile, the health monitoring system is built upon several wireless sensors. As a consequence, the IEEE 802.15.4 is very suitable for it. However, the standard IEEE 802.15.4 protocol cannot provide the guarantee of reliability and real-time [6][7]. The IEEE 802.15.4 protocol has an optional CFP (Contention Free Period) which can provides GTSs (Guaranteed Time Slots) for those special nodes. But the number of available GTSs is very limited in each superframe and the GTS allocation scheme is also quite rough. To address this problem, we attempt to modify the IEEE 802.15.4 to satisfy the requirement of reliable and real-time cyber-physical applications such as health monitoring.

In this paper, a new protocol, i.e. Ada-MAC, based on IEEE 802.15.4 standard is proposed to support the several special requirements: real-time transmission, reliability, collision avoidance and adaption. We implant the time-trigger mechanism, the priority queue mechanism, adaptive mini-time slot allocation strategy in the Ada-MAC protocol. In this way, the protocol could support the requirement of allocating the GTSs adaptively for those special nodes according to the data types. High priority data can obtain a reliable and real-time transmission. Low priority data can be transmitted in other slot time not assigned initially. We simulate the proposed protocol under a health monitoring scene based on the OMNET++ simulation platform. In order to investigate the performance of the proposed protocol, we compare it with the IEEE 802.15.4 CSAM/CA protocol, and the result is satisfactory.

The organization of this paper is as follows. Section II highlights the related work and motivation about our research. In Section III, we present a detailed description of the Ada-MAC protocol. The simulation settings and the simulation results are shown in Section IV and Section V, respectively. At last, we make the conclusions in Section VI.

## II. Related Work And Motivation

IEEE 802.15.4 is a standard designed for low-rate wireless personal area networks. It covers the physical layer and the MAC layer of a low-rate wireless network. It has been applied to a broad range of fields by virtue of its interoperability, low power and cost efficiency. Many researchers have devoted to investigating the performance of IEEE 802.15.4 protocol in different conditions [6-11]. It has been found that the IEEE 802.15.4 protocol has the problem of unreliability and unpredictable (and possibly large) latency.

Giuseppe et al. [11] show a comprehensive analysis of IEEE 802.15.4 protocol in MAC unreliability problem including packet dropout and latency, etc. Park et al. [12] present a novel adaptive MAC algorithm and the stations can adjust the MAC parameter adaptively using the proposed algorithm to guarantee the reliability and delay constrains of

This work is partially supported by Nature Science Foundation of China under grant No. 60903153, the Fundamental Research Funds for Central Universities, the SRF for ROCS, SEM, and DUT Graduate School (JP201006) the IEEE 802.15.4 protocol, according to current transmission condition. In [13], Anis et al. add the priority queue to the IEEE 802.15.4 protocol. In each node, the packet with different priority have different value of the CSMA/CA parameter. In [14], a new priority-based algorithm for the IEEE 802.15.4 beacon-enable network is proposed in order to alleviate an end-to-end delay.

But all these improvements are considered in the slotted CSMA/CA mechanism. In addition to improve in CSMA/CA mechanism, many researchers start to pay attention to the GTS allocation mechanism in the CFP.

In [15], the authors present a new approach to allocate GTSs in the IEEE 802.15.4 protocol which is called i-GAME that allows sharing the same GTSs between multiple flows based on their traffic specifications and delay requirements to guarantee the reliable and timely transmission. An adaptive and real-time GTS allocation mechanism called ART-GAS has been given in [16] which has two stages: Operate a service different mechanism that dynamically assigns data–base priorities and rate-based priorities to all nodes in the first stage, and allocate GTS resources to node according to their priorities assigned in the second stage.

In [17], the authors use the Time-Triggered Communications (TTC) over the IEEE 802.15.4 protocol and give a preliminary solution for the transmission of real-time time-trigger traffic. In [18], Afonso et al. modify the IEEE 802.15.4 protocol and presented a new MAC protocol supporting for real-time and loss intolerant traffic through the contention-free operation, the retransmission scheme flexibility and the high throughput efficiency.

In this paper, we expand the previous research on the IEEE 802.15.4 protocol and propose the Ada-MAC protocol. The main contributions of the paper are: 1) We implant the priority queue and the time-trigger mechanism into the IEEE 802.15.4 protocol. 2) The protocol could support the adaptive GTS allocation mechanism by improving the utilization ratio of GTSs and reducing the waste of time slots. 3) We simulate the proposed protocol by the OMNET++ simulator and evaluate the performance as well. 4) A comparison of the performance between the proposed protocol and the IEEE 802.15.4 CSMA/CA protocol is provided.

## III. ADA-MAC PROTOCOL

The Ada-MAC protocol is a hybrid schedule based on the Time-Triggered Protocol (TTP), and contention based on the CSMA/CA protocol. It satisfies the real-time transmission and the collision avoidance by using the GTSs, and is adaptively supported by adjusting the durations of CFP. The protocol could allocate appropriate numbers of time slots for the specific nodes which have burst or important data based on time-trigger mechanism. At the same time, the other nodes could transmit their data in the remaining time slots following the CSMA/CA mechanism.

### A. Superframe structure

The Ada-MAC protocol is based on beacon-enabled IEEE 802.15.4 MAC standard and is used in a star topology. The superframe structure of the protocol is shown in Fig. 1. The superframe is divided into a fixed number of mini-time slots (Current implementation is 64) and each mini-time slot is long enough to transmit one data. During the communication, the coordinator broadcasts the beacon frames periodicity in which carry the information of the new superframe structure including the position of the mini-time slot pre-allocated to each node that has important data such as the duration of CFP, begin time of CAP, beacon interval, etc. A beacon frame also means a new start of a superframe. When a node receives a beacon frame, it should synchronize with the coordinator first which is a very important step for the time-trigger mechanism.

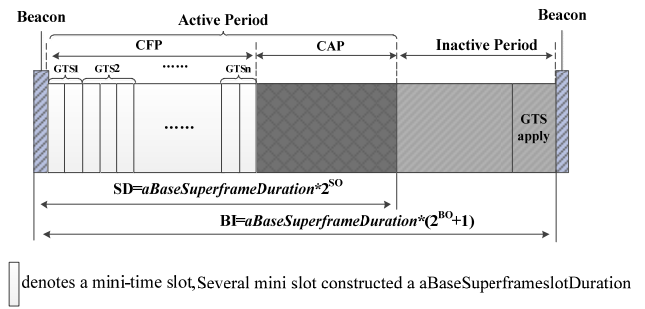

$$SD = aBaseSuperframeDuration*2^{SO}$$
$$BI = aBaseSuperframeDuration*(2^{BO}+1)$$

denotes a mini-time slot, Several mini slot constructed a aBaseSuperframeslotDuration

Figure 1.    Structure of superframe

The superframe can be subdivided into the following three periods: 1) The Contention Free Period (CFP); 2) The Contention Access Period (CAP); 3) The inactive period. Different from the standard IEEE 802.15.4 protocol, we swap the position of CAP and CFP in the superframe, as shown in Fig.1.

The CFP contains a number of GTSs allocated by the PAN coordinator to the specific nodes for sending real-time data. Each GTS may contain one or more mini-time slots and only belongs to one node. The CFP uses the time-triggered mechanism and each node can be triggered at the start of its own GTSs and transmits data according to the GTSs information announced previously in the beacon frame of the current superframe. When its own GTSs end, the node would turn off the transceiver and go to sleep. The GTS assignment for a node is valid only in the current superframe. The transmissions during the CFP can provide the requirements of reliability and less time for the time-critical data. In addition, we remove the restriction of seven GTSs per superframe in the Ada-MAC protocol. Therefore the duration of the CFP can be regulated according to the condition of demands and the max duration of the CFP can be expanded to the whole active period.

The CAP is placed after the CFP. During the CAP, each station can transmit non-real-time data using the slotted CSMA/CA mechanism. The whole operation process of the CAP is the same with the standard IEEE 802.15.4 protocol.

### B. Priority queue mechanism

The proposed Ada-MAC protocol defines three data types: burst data, periodic data and normal data. The burst data and periodic data require real-time and reliable transmission while the normal data do not need specified requirement, as shown in Table I. Here, we refer to the real-time as timely treatment of data with a deadline and no low-delay communication. The Priority Queue mechanism allocates separate queues to different types of data. The packets within each queue are maintained in a FIFO order. The Priority Queue mechanism can reduce the queuing delays of the high priority data.

TABLE I.    PRIORITY OF DATA

| Data type | Priority | Requirement |
|---|---|---|
| Burst Data | Highest | Real-time |
| Periodic Data | High | Real-time |
| Normal Data | Low | Non-real-time |

In this protocol, we fix the number of queues to three, as shown in Fig. 2. When the data frames arrive, the queue system at first classifies the frames based on the frame type identified by the upper layer and stores them into corresponding queues. The packets in low-priority queue can only be transmitted in the CAP using the CSMA/CA mechanism. During the CFP, when the burst data and periodic data are all waiting for transmission, the mac will choose the burst data to be transmitted first.

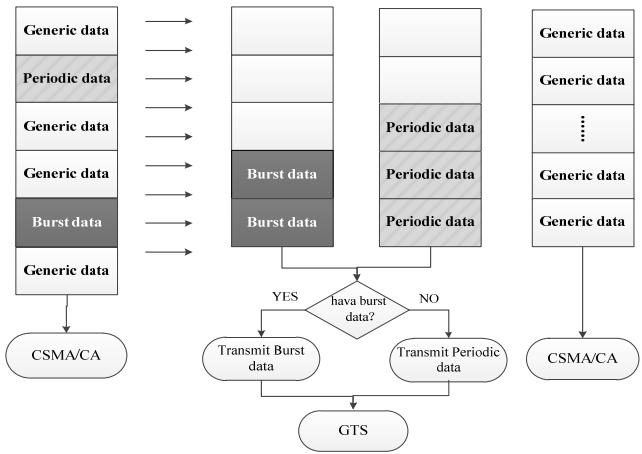

Figure 2.   Priority queue

### C.  Adaptive mini-time slot assignment strategy

In the previous section, we mentioned that we divide the whole superframe duration into 64 mini-time slots and remove the restriction of seven GTSs in each superframe in the Ada-MAC protocol. In this way, more nodes can obtain the GTSs to transmit their real-time data frames in a reliable and timeless way.

In order to reduce the number of wasted time slots and achieve dynamic distribution of resources, all nodes will send GTS allocation requests to the PAN coordinator at the end of each superframe. Before sending the request, the node will check the number of the burst data and periodic data waiting for transmission in the priority queues and record this information in the GTS allocation request if it has. After the coordinator receives the GTS request, it schedules the structure of the CFP by using the Mini-time slot assignment strategy. The coordinator gives priority to assign the GTSs for the nodes that have burst data to transmit. And the nodes which have more burst data would receive the more forward allocated GTSs. If the nodes have the same number of burst data, the one that has more periodic data will come first. The mini-time slot assignment strategy has been given by Algorithm 1. At last, the schedule of the CFP is conveyed in the beacon frame by the field "GTS Allocation List", as shown in Fig. 3.

Algorithm 1. Mini-time slot assignment strategy

---

1: $startslot = 1$, $j = 0$;
2: set maxslot // the max number
3: GR= {$length$, $MacAddress$, $brust$, $periodic$}
4: GL= {$startlot$, $length$, $MacAddress$}
5: **while** GR =! NULL **do**
6:     $index = 0$;
7:     N=length of GR
8:     **for** $i = 0,1,2,…,N-1$ **do** //N means the length of GR
9:         **if** GR[$i$].$brust$ > GR[$index$].$brust$ **then**
10:             $index = i$;
11:         **end if**
12:         **if** GR[$i$].$brust$ == GR[$index$].$brust$ **then**
13:             **if** GR[$i$].$periodic$ > GR[$index$].$periodic$ **then**
14:                 $index = i$;
15:             **end if**
16:         **end if**
17:     **end for**
18:     **if** $startslot$ < $maxslot$ **then**
19:         Assign $length$ muni-slots to the slave node and start in $startslot$
20:         GL[$j$].startslot=$startslot$,
21:         GL[$j$].$length$=G[index].$length$
22:         GL[$j$].$Macaddress$=G[$index$].$Macaddress$
23:         delete G[$i$] from G;
24:         $j = j+1$;
25:         $startslot$ = $startslot$+GL[$j$].length;
26:     **end if**
27:**end while**

---

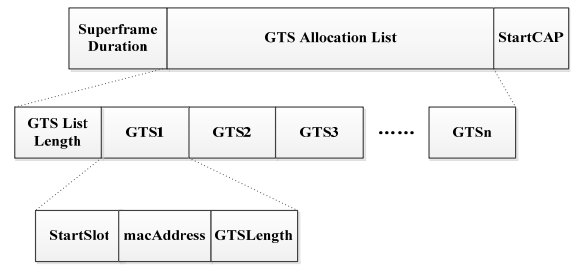

Figure 3.   Structure of GTS allocation list in beacon frame

In the Algorithm 1, GR stands for the GTS request sent by the nodes, and it contains four parameters: length, MacAddress, burst and periodic. Length signifies the number of mini-time slot requested. MacAddress signifies the mac address of the request node. Burst denotes the number of burst data that the node needs to transmit; periodic denotes the number of periodic data that node needs to transmit. GL stands for the GTS List scheduled by the PAN coordinator. It contains three parameters: startslot, length and MacAddress. Startslot denotes the GTS starting mini-time slots allocated by the PAN coordinator for the applying nodes. 0 means no GTSs for the node. Length denotes the number of mini-time slots for the GTSs.

## IV. SIMULATION SETTINGS

In order to evaluate the performance of the Ada-MAC protocol, we apply the proposed protocol to the health monitoring application composed of a base node (the PAN coordinator) and several wireless monitoring nodes, as presented in Fig. 4. The PAN coordinator collects the data from the sensor nodes deployed in different parts of the body. These sensor nodes can sense physiological signals such as electrocardiogram (ECG), blood pressure, temperature and accelerometer for the heath care monitoring service. They send data to the PAN coordinator periodically. Some of them need to be delivered correctly within a predefined deadline. However, emergency data may be generated randomly and need to be transmitted in time. The max delay bound of each data type we assumed is presented in Table II.

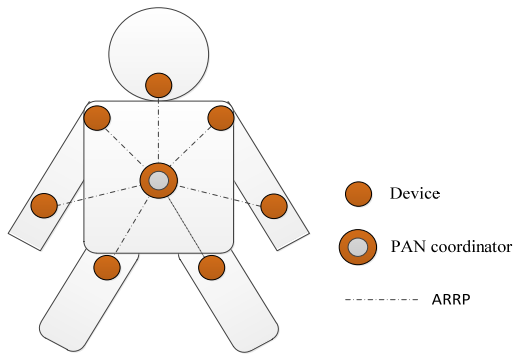

Figure 4. Health monitoring system

TABLE II. MAX DELAY BOUND OF DATA

| Data type | Max delay bound |
|---|---|
| Burst data | 250ms |
| Periodic data | 450ms |
| Normal data | --- |

We simulated the health monitoring scene based on the Ada-MAC protocol using the OMNET++ simulator. In the simulation, a star topology with a single PAN coordinator and 16 nodes deployed in the area of 100cm *100cm is considered. The nodes are uniformly distributed around a 50 centimeters radius circle while the PAN coordinator is placed at the center of the circle. Each transmission is a single-hop. There is no hidden node in the scene. All the nodes are set to be in each other's radio range and each node can learn the other nodes easily. The detailed simulation settings are shown in Table III.

TABLE III. SIMULATION PARAMETERS

| Parameter | Value |
|---|---|
| Carrier Frequency | 2.4 GHz |
| Transmitter power | 1 mW |
| Carrier sense sensitivity | -85 dBm |
| Bitrate | 250 Kbps |
| Queue length | 10 packets |
| Traffic type | exponential |
| Run time | 2000s |
| MaxBE | 5 |
| MinBE | 3 |
| MaxNB | 4 |
| MaxFrameRetries | 3 |
| MAC payload size (MSDU size) | 50byte |
| Superfame order (SO) | 4 |
| Beacon order (BO) | 4 |
| Number of end devices | 16 |

By changing the packet generation rate of different types of data, we intend to investigate the performance of the Ada-MAC protocol including the timeliness, reliability, and resource efficiency. In order to meet these requirements, we select mean (max) end-to-end delay, packet loss rate and packet delivery ratio as performance metrics.

## V. SIMULATION RESULTS

This section will present the performance of the Ada-MAC protocol and compare it with the performance of the IEEE 802.15.4 CSMA/CA protocol. In this way, we can analyze the advantages and disadvantages of the proposed protocol easily.

We consider six different scenarios presented in Table IV. Here we assume that the burst data is generated randomly. The probability of the burst data appears to be $P_{burst}$= 0.5‰, 1‰, 2‰ under the two different protocols. The generation interval of the periodic data ranges from 0.1s to 0.7s in each scenario. The generation interval of the normal data is fixed to 0.05s.

Fig. 5 and Fig. 6 clearly show the mean end-to-end delay of the periodic data and the burst data for all scenarios. We can find that the periodic data in the IEEE 802.15.4 protocol have a very high average latency. On the contrary, the periodic data always have a very low average latency in the Ada-MAC protocol. We find that with the increasing of the generation interval of the periodic data, the average latency decline slowly. The average delay of the periodic data and the burst data rise as the value of $P_{burst}$ becomes larger, especially in the smaller generation interval of the periodic data.

| Scenario | Commutation protocol | $P_{burst}$ |
|---|---|---|
| Scenario 1 | Ada-MAC protocol | 2‰ |
| Scenario 2 | Ada-MAC protocol | 1‰ |
| Scenario 3 | Ada-MAC protocol | 0.5‰ |
| Scenario 4 | IEEE 802.15.4 CSMA/CA protocol | 2‰ |
| Scenario 5 | IEEE 802.15.4 CSMA/CA protocol | 1‰ |
| Scenario 6 | IEEE 802.15.4 CSMA/CA protocol | 0.5‰ |

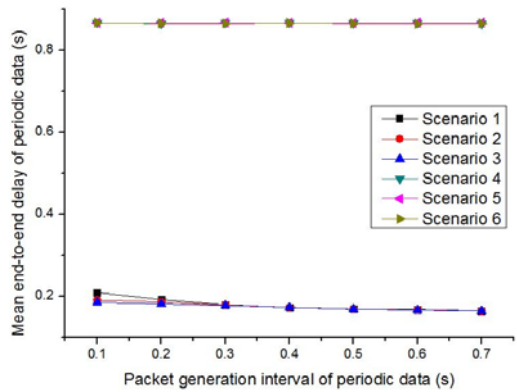

Figure 5.   Mean end-to-end delay of periodic data

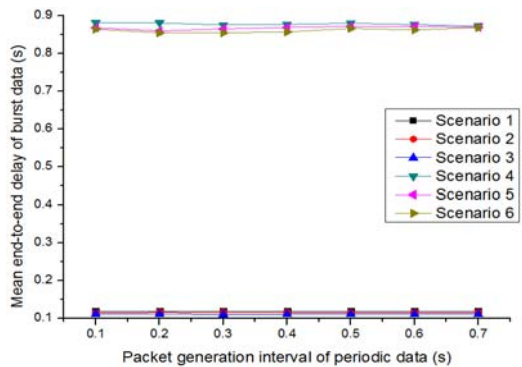

Figure 6.   Mean end-to-end delay of burst data

The max end-to-end delay of the burst data and periodic data are depicted in Fig. 7 and Fig. 8 respectively. In the Ada-MAC protocol, the periodic data experience the max delay within 420s and the burst data within 260ms. Compared with the max delay of the real-time data in the IEEE 802.15.4 protocol, the Ada-MAC protocol provides an acceptable delay for the real-time data (including the periodic data and burst data).

According to the four figures above, we can learn that in the IEEE 802.15.4 protocol, the buffered packet bring up the severe contentions and cause the long delay of the real-time data. However, the priority queue adopted in the Ada-MAC protocol can make sure that the high priority data be delivered to their destination in a bounded time interval by the GTSs during the CFP.

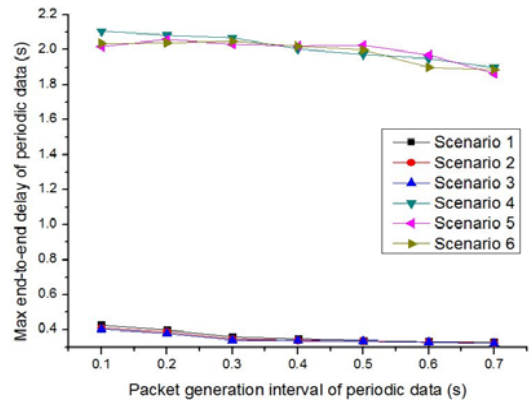

Figure 7.   Max end-to-end delay of periodic data

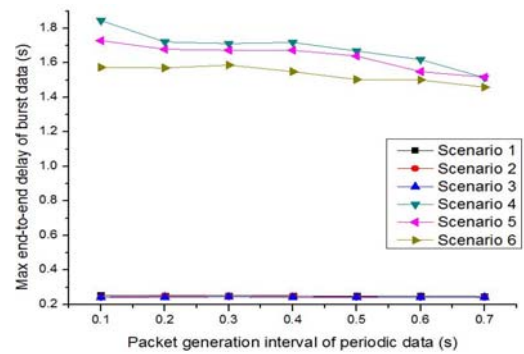

Figure 8.   Max end-to-end delay of burst data

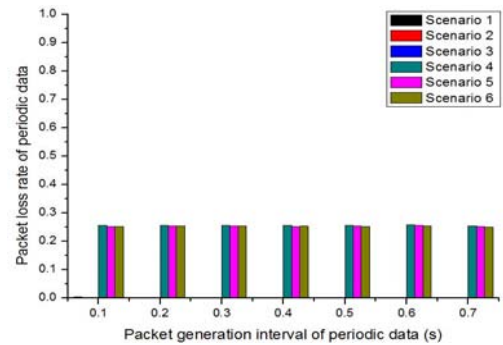

Figure 9.   Packet loss rate of periodic data

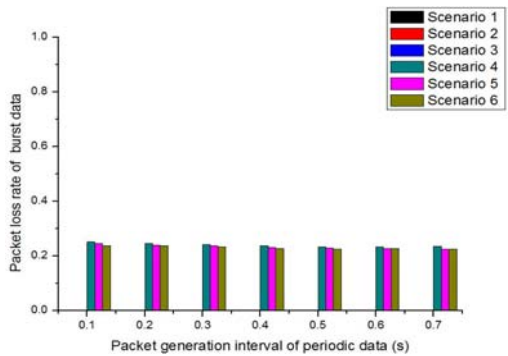

Figure 10. Packet loss rate of burst data

In Fig. 9 and Fig. 10, the packet loss rates of the real-time data are compared with respect to two different protocols. From the two figures, we can see that the packet loss rate in the IEEE 802.15.4 CSMA/CA protocol stays at about 24%. But in the Ada-MAC protocol, the packet loss rate is nearly 0. This is because that the high priority data can be transmitted adaptively during the CFP in the proposed protocol by avoiding the conflict during the transmission.

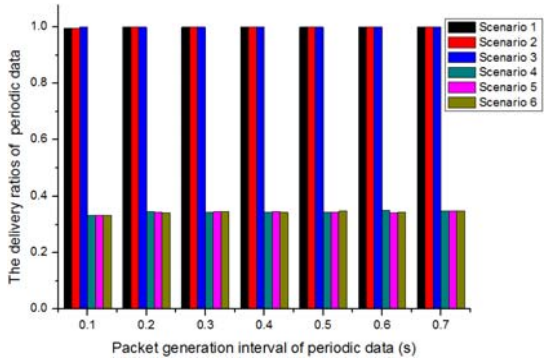

Figure 11. Delivery ratio of periodic data

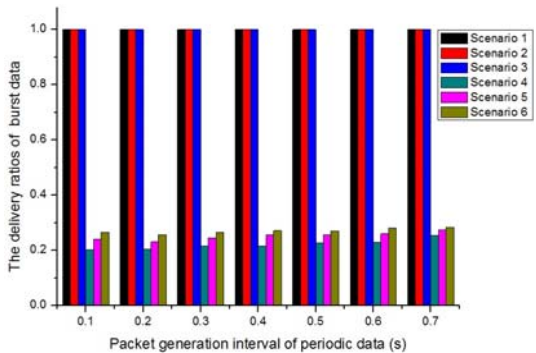

Figure 12. Delivery ratio of burst data

In Fig. 11 and Fig. 12, the delivery ratio of the real-time data is compared in the two protocols. The results show that the delivery ratio in the IEEE 802.15.4 protocol is very low only 20%, which represents a deep effect on the reliability of the protocol. On the contrary, the delivery ratio in Ada-MAC protocol always keeps in 100% with the generation interval of periodic data increasing. The reason is that the Ada-MAC protocol uses the priority queue and mini-time slot allocation strategy to provide the reliability and timeless transmission for the real-time data.

## VI. CONCLUSIONS

This paper has presented the Ada-MAC protocol for cyber-physical systems, taking healthcare monitoring as an example application. This protocol can be used in different transmission mode according to different data types and can provide guaranteed time slots adaptively during the CFP for the nodes which have important packets. With the mechanism of the priority queue, the nodes can learn whether they have real-time data to transmit or not and apply GTS to the coordinator. Then the coordinator uses the mini-time slot allocation strategy to allocate the GTSs for the nodes adaptively. We have implemented the protocol and evaluated its performance as compared against the traditional IEEE 802.15.4 CSMA/CA protocol. The result shows that the proposed protocol performs better in guaranteeing the reliability and timeless transmission for real-time data. In the future, we will continue to test the performance of the protocol in different situations.